\documentclass[aps,prd,preprint,nofootinbib]{revtex4}
\usepackage{epsf,graphicx,amssymb,amsfonts,amsmath}

\begin{document}
\title{Renormalization of Loop Functions in QCD}
\author{Matthias Berwein, Nora Brambilla  and Antonio Vairo}
\affiliation{Physik-Department, Technische Universit\"{a}t M\"{u}nchen, James-Franck-Str.~1, 85748 Garching, Germany}

\begin{abstract}
We give a short overview of the renormalization properties of rectangular Wilson loops, the Polyakov loop correlator and the cyclic Wilson loop. We then discuss how to renormalize loops with more than one intersection, using the simplest non-trivial case as an illustrative example. Our findings expand on previous treatments. The generalized exponentiation theorem is applied to the Polyakov loop correlator and used to renormalize linear divergences in the cyclic Wilson loop.
\end{abstract}

\maketitle

\newpage

\section{Introduction}
We will discuss some loop functions with regard to their renormalization properties. By \textit{loop functions} we mean the vacuum or thermal expectation value of a number of Wilson lines in an $SU(N_c)$ gauge theory, which are closed and traced and each trace is normalized by the number of colours $N_c$.

Rectangular Wilson loops are of special interest, because they are related to the quarkonium potential, and therefore they have been studied in detail~\cite{RWL1,RWL2,RWL3,RWL4,RWL5}. They consist of four straight Wilson lines, two along the time direction at fixed positions in space and at relative distance $\mathbf{r}$, and two along the direction of $\mathbf{r}$ at fixed times and at temporal separation $t$.

It is known that rectangular Wilson loops are UV divergent even after charge renormalization~\cite{Ren1,CuspDiv}. They need to be renormalized by a multiplicative constant, which is given in the $\overline{\mathrm{MS}}$-scheme by $\displaystyle Z_{4c}=\exp\left[-\frac{2C_F\,\alpha_s}{\pi\bar{\varepsilon}}+{\cal O}(\alpha_s^2)\right]$, where~$C_F$ is the quadratic Casimir of the fundamental representation, the space-time dimension $D=4-2\varepsilon$ and $1/\bar{\varepsilon}=1/\varepsilon-\gamma_E+\ln4\pi$. This additional divergence comes from the four corners of the Wilson loop, where the contour has cusps of angle $\pi/2$.

At finite temperature the Polyakov loop correlator $P_c(r,T)$ plays a role similar to the rectangular Wilson loop in the vacuum. It is related to the free energy of a static quark-antiquark pair~\cite{PLC1,PLC2}. Polyakov loops are Wilson lines spanning the whole of the imaginary time direction from $\tau=0$ to $\tau=\beta=1/T$ at fixed spatial position. They are closed loops because of the periodic boundary conditions of the imaginary time formalism. The Polyakov loop correlator consists of two traced Polyakov loops at spatial distance $\mathbf{r}$. This quantity is free of UV divergences in dimensional regularization after charge renormalization.

In our publication~\cite{RCWL} we have studied the renormalization properties of the cyclic Wilson loop $W_c(r,T)$, which is closely related to the two loop functions described above. It is a rectangular Wilson loop at finite temperature where the temporal Wilson lines are given by Polyakov loops. However, its divergence structure does not match that of a vacuum rectangular Wilson loop~\cite{Laine}, because renormalization with a multiplicative constant fails.

The reason for this behaviour lies in the periodic boundary conditions of the imaginary time formalism. The corners of the cyclic Wilson loop lie at $(0,\pm\mathbf{r}/2)$ and $(\beta,\pm\mathbf{r}/2)$, but because $\tau=0$ and $\tau=\beta$ are identified, these points are identical and should therefore be treated as intersections instead of cusps. The renormalization properties of loop functions with cusps and intersections have been studied in general in~\cite{Ren1,Ren2}. Whenever a loop has points of self intersection, renormalization mixes it with other associated loop functions, which have identical contours except for a different path ordering prescription at the intersection. In this case the cyclic Wilson loop mixes with the Polyakov loop correlator. If one diagonalizes the mixing matrix, one obtains multiplicatively renormalizable quantities, which are here given by $P_c$ itself and the difference $W_c-P_c$. The renormalization constant for the latter is given in dimensional regularisation by $Z_{\,W_c-P_c}=\displaystyle\exp\left[-\frac{C_A\,\alpha_s}{\pi\bar{\varepsilon}}+{\cal O}(\alpha_s^2)\right]$, where~$C_A$ is the quadratic Casimir of the adjoint representation.

\section{Loop functions with more than one intersection}\label{sec1}
We would like to point out here that the treatment of multiple intersections in~\cite{Ren2}, on which the argument of~\cite{RCWL} was based, seems not to rely on the most general assumptions. Their statement is that for each intersection point there will be a renormalization matrix which depends only on the angles between the Wilson lines at the intersection. Let us look at the simplest case of loop functions with multiple intersections, i.e.~loop functions with two intersections with two incoming and two outgoing lines each. We can distinguish two different situations: the two intersection points are connected by either two~$(I)$ or four Wilson lines~$(II)$.

\begin{figure}[t]
 \includegraphics[width=\linewidth]{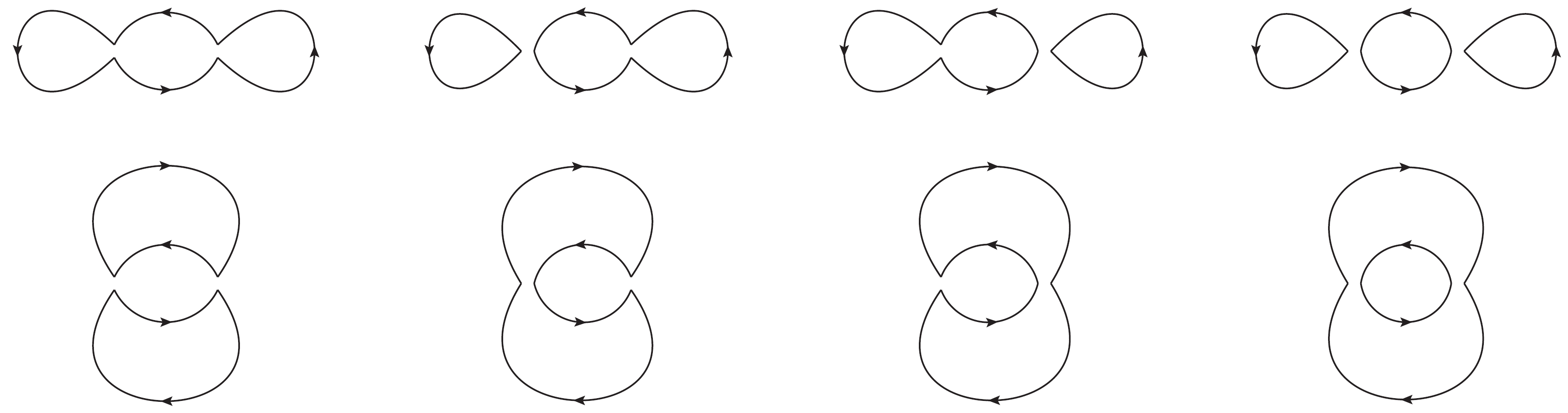}
 \caption{All possible contours for situations $(I)$ and $(II)$. From left to right, the first row corresponds to loop functions $L^{(I)}_{11}$, $L^{(I)}_{21}$, $L^{(I)}_{12}$, $L^{(I)}_{22}$ and the second row corresponds to $L^{(II)}_{11}$, $L^{(II)}_{21}$, $L^{(II)}_{12}$, $L^{(II)}_{22}$. The contours are drawn apart at the intersections in order to show how the Wilson lines are connected, nevertheless they should be understood to touch at those points.}
 \label{contours}
\end{figure}

In either case there are four different Wilson lines starting and ending at an intersection point. The associated loop functions which mix under renormalization are given by all possibilities in which the colour indices of incoming Wilson lines can be contracted to those of outgoing Wilson lines at each intersection. In this case there are two possibilities at each intersection to combine the colour indices, so there are in total four associated loop functions. We will denote them as $L^{(I)}_{ij}$ and $L^{(II)}_{ij}$, where the indices $i$ and $j$ label the path ordering prescriptions at the first and second intersection respectively. The details of these definitions are illustrated schematically in figure~\ref{contours}.

\begin{figure}[t]
 \centering
 \includegraphics[width=\linewidth]{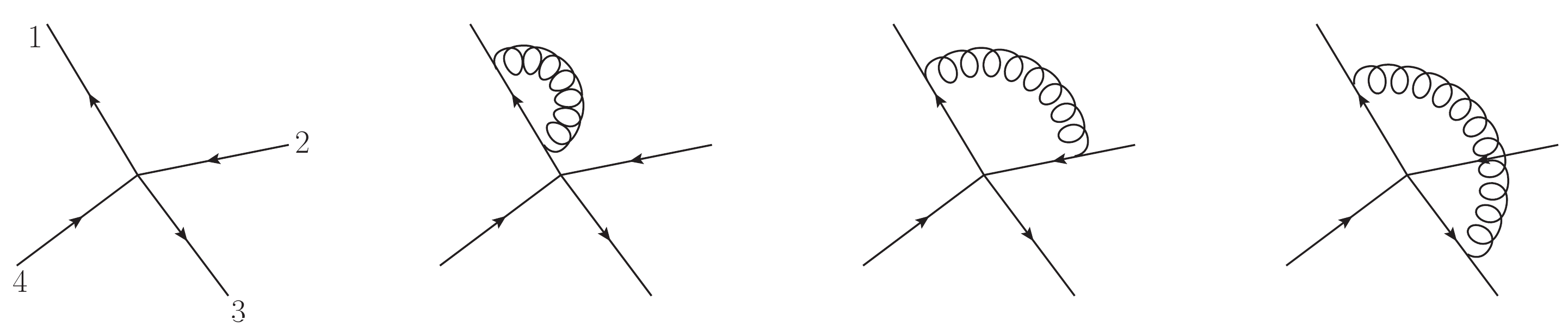}
 \caption{The labels for the intersection angles are illustrated in the left picture. The second figure shows a contribution to the intersection divergence which does not depend on the angles. The two figures on the right show contributions depending on the angles $\alpha_{12}$ and $\alpha_{13}$ respectively. The divergence of the rightmost diagram gets an additional minus compared to a corresponding cusp divergence, because it involves two outgoing Wilson lines.}
 \label{angles}
\end{figure}

The important point is that both $(I)$ and $(II)$ loop functions should be renormalized by the same matrices, provided that the angles at the intersections are the same. We will check this at leading order. The angles will be called $\alpha_{kl}$ for the left intersection and $\beta_{kl}$ for the right intersection, where the indices $k$ and $l$ label the Wilson lines which define the angle according to figure~\ref{angles}. In general, an intersection of four Wilson lines has six different angles, which in four space-time dimensions are all independent.

When there is only one gluon, the divergences at the intersections come from the same diagrams as in the case of cusp divergences. So we can use the general result for cusp divergences in dimensional regularisation obtained in~\cite{CuspDiv}:
\begin{equation}
 \frac{C_F\,\alpha_s}{2\pi\varepsilon}\left(1+(\pi-\gamma)\cot\gamma\right)=\Delta_0+\Delta(\gamma)\,.
\end{equation}

This consists of two parts, $\Delta(\gamma)$ depends on the cusp angle $\gamma$ and $\Delta_0$ is independent of~$\gamma$. $\Delta_0$ comes from diagrams where the gluon starts and ends at the same Wilson line close to the intersection. In the case of a cusp there are two such diagrams, for the intersections there are four, so the angle-independent divergence of each intersection is twice that of a cusp.

The angle-dependent part $\Delta(\gamma)$ comes from diagrams where the gluon spans across an angle close to the intersection. In the case of a cusp there is always one incoming and one outgoing line, at an intersection there can be two incoming or two outgoing lines, in which case the divergence is minus that of a cusp of the same angle. Illustrations for this can also be found in figure~\ref{angles}.

It is determined by the contours shown in figure~\ref{contours}, on which angles the intersection divergences depend. The contours consist of either one, two or three closed Wilson loops. The trace of a single colour matrix vanishes, so diagrams where the gluon connects two different Wilson loops do not contribute. With this in mind we obtain the following divergences for the different loop functions:
\begin{align}
 \mathrm{Div}(L^{(I)}_{11})=&\,4\Delta_0+\Delta(\alpha_{12},\alpha_{23},\alpha_{34},\alpha_{14},\beta_{12},\beta_{23},\beta_{34},\beta_{14})-\Delta(\alpha_{13},\alpha_{24},\beta_{13},\beta_{24})\,,\label{Div1}\\
 \mathrm{Div}(L^{(I)}_{21})=&\,4\Delta_0+\Delta(\alpha_{23},\alpha_{14},\beta_{12},\beta_{23},\beta_{34},\beta_{14})-\Delta(\beta_{13},\beta_{24})\,,\label{Div2}\\
 \mathrm{Div}(L^{(I)}_{12})=&\,4\Delta_0+\Delta(\alpha_{12},\alpha_{23},\alpha_{34},\alpha_{14},\beta_{23},\beta_{14})-\Delta(\alpha_{13},\alpha_{24})\,,\label{Div3}\\
 \mathrm{Div}(L^{(I)}_{22})=&\,4\Delta_0+\Delta(\alpha_{23},\alpha_{14},\beta_{23},\beta_{14})\,,\label{Div4}\\
 \mathrm{Div}(L^{(II)}_{11})=&\,4\Delta_0+\Delta(\alpha_{12},\alpha_{34},\beta_{12},\beta_{34})\,,\label{Div5}\\
 \mathrm{Div}(L^{(II)}_{21})=&\,4\Delta_0+\Delta(\alpha_{12},\alpha_{23},\alpha_{34},\alpha_{14},\beta_{12},\beta_{23},\beta_{34},\beta_{14})-\Delta(\alpha_{13},\alpha_{24},\beta_{13},\beta_{24})\,,\label{Div6}\\
 \mathrm{Div}(L^{(II)}_{12})=&\,4\Delta_0+\Delta(\alpha_{12},\alpha_{23},\alpha_{34},\alpha_{14},\beta_{12},\beta_{23},\beta_{34},\beta_{14})-\Delta(\alpha_{13},\alpha_{24},\beta_{13},\beta_{24})\,,\label{Div7}\\
 \mathrm{Div}(L^{(II)}_{22})=&\,4\Delta_0+\Delta(\alpha_{23},\alpha_{14},\beta_{23},\beta_{14})\,,\label{Div8}
\end{align}
where we have introduced the shorthand notation $\Delta(\gamma_1,\gamma_2,\dots)=\Delta(\gamma_1)+\Delta(\gamma_2)+\dots$.

At zeroth order in $\alpha_s$ all loop functions are equal to $1$, therefore also the renormalization matrices should be given by a unit matrix. We can then make a general ansatz for the renormalization matrices $Z^{(1)}$ and $Z^{(2)}$ associated with the left and right intersection respectively:
\begin{equation}
 Z^{(1)}=\left(\begin{array}{cc} 1-a & b \\ c & 1-d \end{array}\right)\,,\hspace{20pt}Z^{(2)}=\left(\begin{array}{cc} 1-A & B \\ C & 1-D \end{array}\right)\,,
\end{equation}
where all variables are of order $\alpha_s$.

If we now require $Z^{(1)}_{ik}Z^{(2)}_{jl}L^{(I)}_{kl}$ and $Z^{(1)}_{ik}Z^{(2)}_{jl}L^{(II)}_{kl}$ to be finite at leading order in $\alpha_s$, we arrive at the following system of equations:
\begin{align}
 (a-b)+(A-B)&=\mathrm{Div}(L^{(I)}_{11})\,, & (a-b)+(A-B)&=\mathrm{Div}(L^{(II)}_{11})\,,\label{I}\\
 (d-c)+(A-B)&=\mathrm{Div}(L^{(I)}_{21})\,, & (d-c)+(A-B)&=\mathrm{Div}(L^{(II)}_{21})\,,\label{II}\\
 (a-b)+(D-C)&=\mathrm{Div}(L^{(I)}_{12})\,, & (a-b)+(D-C)&=\mathrm{Div}(L^{(II)}_{12})\,,\label{III}\\
 (d-c)+(D-C)&=\mathrm{Div}(L^{(I)}_{22})\,, & (d-c)+(D-C)&=\mathrm{Div}(L^{(II)}_{22})\,.\label{IV}
\end{align}
These four equations are not independent, if we look at the left-hand side, we see that $\eqref{I}+\eqref{IV}=\eqref{II}+\eqref{III}$ and the same should be true for the right hand side. By comparing with equations~\eqref{Div1}~-~\eqref{Div8} we see that this is indeed the case for loop functions of type~$(I)$ but in general not for type~$(II)$. We therefore conclude that type~$(II)$ loop functions cannot be renormalized by two independent renormalization matrices for each intersection.

In the case~$(I)$ there is no such problem and the equations can be solved. We have a system of three independent equations for four independent variables: $(a-b)$, $(d-c)$, $(A-B)$ and $(D-C)$. The solution can be made unique by the additional requirement that $Z^{(1)}$ can only depend on the angles~$\alpha_{ij}$, $Z^{(2)}$ only on~$\beta_{ij}$ and both should have the same form. We get
\begin{align}
 (a-b)&=2\Delta_0+\Delta(\alpha_{12},\alpha_{23},\alpha_{34},\alpha_{14})-\Delta(\alpha_{13},\alpha_{24}), & (d-c)&=2\Delta_0+\Delta(\alpha_{23},\alpha_{14}),\\
 (A-B)&=2\Delta_0+\Delta(\beta_{12},\beta_{23},\beta_{34},\beta_{14})-\Delta(\beta_{13},\beta_{24}), & (D-C)&=2\Delta_0+\Delta(\beta_{23},\beta_{14}),
\end{align}
and it is easy to check that $Z^{(1)}$ and $Z^{(2)}$ are the same as the leading order renormalization matrices for loop functions with just one intersection with angles $\alpha_{ij}$ or $\beta_{ij}$.

It seems that the statements on multiple intersections in reference~\cite{Ren2} implicitly assume that the divergence structures at each intersection have no influence on each other. 
But this is only true in case~$(I)$. There are two things which determine the divergence structure at an intersection. The first is the path ordering prescription at the intersections, which is determined by the indices~$i$ and~$j$. The second is how the outgoing lines are connected to incoming lines away from the intersection. In our example, line~$1$ is always connected to line~$4$ and line~$3$ always to line~$2$ at both intersections for the type~$(I)$ loop functions, where we use the same labels that have been introduced for the angles in figure~\ref{angles}. But for type~$(II)$ at each intersection the lines~$1$ and~$3$ can be connected to lines~$2$ or~$4$ depending on the other intersection.

However, if we drop the assumption that both intersections can be renormalized independently and instead of two indices for both intersections we use one index labelling all four loop functions, then the whole line of argument of~\cite{Ren2} can be repeated with only slight adjustments. This means that the type~$(II)$ loop functions are renormalized by a $4\times4$~matrix that in contrast to case $(I)$ is not given by the tensor
product of two $2\times2$~matrices.

Therefore we expect the most general renormalization prescription for loop functions to read like this: there is one renormalization constant for each cusp and one renormalization matrix for each independent set of intersections. A set of intersections is called independent, if any other set is connected to it through at most two Wilson lines. The reason for this classification is that if there are only one outgoing and one incoming line leading from one set to the other, then there is only one way in which those can be connected. As soon as there are more possibilities to connect outgoing and incoming lines, then each one of them will be realized for some path ordering prescription and we have a situation like in case~$(II)$ of our example.

The renormalization matrices for each independent set depend only on the renormalization scheme, on the angles at the intersections and on the way in which the intersections are connected, but are otherwise completely independent of any specifics of the contour. The intersections of the cyclic Wilson loop are independent from each other, so the ansatz in~\cite{RCWL} with two renormalization matrices for the intersections was justified.

\section{Linear divergences in the cyclic Wilson loop}
Something that was not considered in~\cite{RCWL} is the cancellation of linear divergences in the cyclic Wilson loop. In general, loop functions have power law divergences, which factorize and exponentiate to give a factor $\exp\left[\Lambda L(C)\right]$, where $L(C)$ is the length of the contour and $\Lambda$ is some linearly divergent constant~\cite{Smooth}. In dimensional regularization such linear divergences are absent, so they were not considered in~\cite{RCWL}, but here we would like to show how they cancel in other regularization schemes such as e.g.~lattice regularization.

The cyclic Wilson loop is special in that the two spatial Wilson lines occupy exactly the same points in Euclidean space-time, but they have opposite orientation and are therefore inverse to each other. They do not cancel only because through path ordering they are separated by the Polyakov loops. However, because of this many but not all of the diagrams that would normally contribute to the linear divergence cancel. This means that in the case of the cyclic Wilson loop the cancellation of these linear divergences will be more complicated than for loop functions without such overlapping Wilson lines, where it is sufficient to introduce the renormalization constant $\exp\left[-\Lambda R(C)\right]$.

We will show that the mixing with the Polyakov loop correlator, which we have introduced to remove the intersection divergences, also takes care of the linear divergences. The result of this mixing is that $W_c-P_c$ is multiplicatively renormalizable. It is possible to express this quantity as one loop function. If we use $P(\mathbf{r})$ and $S(\mathbf{r})$ to denote the Polyakov loops and the spatial Wilson lines respectively, then we can write
\begin{align}
 W_c(\mathbf{r})&=\frac{1}{N_c}\left\langle\mathrm{Tr}\left[S(\mathbf{r})P^\dagger(0)S^\dagger(\mathbf{r})P(\mathbf{r})\right]\right\rangle\,,\\
 P_c(\mathbf{r})&=\frac{1}{N^2_c}\left\langle\mathrm{Tr}\left[P^\dagger(0)\right]\mathrm{Tr}\bigl[P(\mathbf{r})\bigr]\right\rangle\,.
\end{align}
The untraced Polyakov loops $P(\mathbf{r})$ are complex $N_c\times N_c$ matrices, which can be decomposed as
\begin{equation}
 P(\mathbf{r})=\frac{1}{N_c}\mathrm{Tr}\bigl[P(\mathbf{r})\bigr]\mathbf{1}_{N_c}+\frac{1}{T_F}\mathrm{Tr}\bigl[P(\mathbf{r})T^a\bigr]T^a\,,
\end{equation}
because the unit matrix $\mathbf{1}_{N_c}$ and the fundamental colour matrices $T^a$, which are normalized as $\mathrm{Tr}\bigl[T^aT^b\bigr]=T_F\delta^{ab}$, form a complete set of linearly independent $N_c\times N_c$ matrices. If we insert this into the definition of $W_c-P_c$ we get
\begin{align}
 W_c(\mathbf{r})-P_c(\mathbf{r})=&\,\frac{1}{N_cT_F}\left\langle\mathrm{Tr}\bigl[P(\mathbf{r})T^a\bigr]\mathrm{Tr}\left[S(\mathbf{r})P^\dagger(0)S^\dagger(\mathbf{r})T^a\right]\right\rangle\notag\\
 &+\frac{1}{N^2_c}\left\langle\mathrm{Tr}\bigl[P(\mathbf{r})\bigr]\mathrm{Tr}\left[S(\mathbf{r})P^\dagger(0)S^\dagger(\mathbf{r})\right]\right\rangle-P_c(\mathbf{r})\notag\\
 =&\,\frac{1}{N_cT_F}\left\langle\mathrm{Tr}\bigl[P(\mathbf{r})T^a\bigr]S^{ab}_A(\mathbf{r})\mathrm{Tr}\bigl[P^\dagger(0)T^b\bigr]\right\rangle\notag\\
 =&\,\frac{T^a_{ji}T^b_{lk}}{N_cT_F}\left\langle P_{ij}(\mathbf{r})S^{ab}_A(\mathbf{r})P^\dagger_{kl}(0)\right\rangle\,,
\end{align}
where now we have one spatial Wilson line $S_A(\mathbf{r})$ in the adjoint representation instead of the two Wilson lines $S(\mathbf{r})$ and $S^\dagger(\mathbf{r})$ in the fundamental representation.

In the last line we have written everything in components. The reason for doing this is that there exists an exponentiation formula for untraced Wilson lines in general colour representations. It has been known for a long time~\cite{Exp1,Exp2} that expectation values of a closed Wilson line can be exponentiated, i.e.~they can be expressed as an exponential of a series of Feynman diagrams. These diagrams are the same as those that would appear in a straightforward perturbative calculation of the loop functions, but there appear less of them in the exponent and they have changed colour coefficients. Recently, this exponentiation property has been generalized in~\cite{Exp3,Exp4}. Their framework is multiparton scattering amplitudes, but the formalism is general.

This generalized exponentiation is to be understood in the following way. Feynman diagrams for untraced Wilson lines have a number of initial and final colour indices, where \textit{initial} and \textit{final} are used in the context of the path ordering of the Wilson lines. A multiplication of two diagrams can be defined as the contraction of the initial indices of the first diagram with the final indices of the other, and an exponential of diagrams is then defined in the usual way as a power series with respect to this multiplication. We will outline here the basic ideas which lead to this exponentiation of untraced Wilson lines, for more details we refer to~\cite{Exp3,Exp4}.

The method is known as the replica trick. Suppose that~$W$ stands for a number of untraced Wilson lines, and~$N$ for some integer. We can expand
\begin{equation}
 \langle W\rangle^N=1+N\ln\langle W\rangle+{\cal O}(N^2)\,,\label{replica}
\end{equation}
so if we want to write~$\langle W\rangle$ as an exponential, then its exponent will be given by the term linear in~$N$ in the expansion of~$\langle W\rangle^N$.

Suppose now that we work in a theory with~$N$ copies of QCD, which do not interact with each other. Instead of calculating~$\langle W\rangle^N$ as the product of~$N$ expectation values of~$W$, we can also calculate it as the expectation value of the product of~$N$ copies of~$W$, provided that they do not interact with each other:
\begin{equation}
 \langle W\rangle^N=\langle W\rangle\cdot\langle W\rangle\cdots\langle W\rangle=\langle W_1\cdot W_2\cdots W_N\rangle\,,
\end{equation}
where the~$W_i$ are defined in the~$i$th copy of QCD in the replicated theory. This last expression introduces a new path ordering: in addition to the path ordering along the contours of the Wilson lines for each~$W_i$, gluons belonging to different copies of QCD will be arranged such that all gluons from~$W_i$ stand on the left of the gluons from~$W_j$ for all~$i<j$.

A Feynman diagram~$D$ for expectation values of Wilson lines can be split into two parts: a kinematical part~$F(D)$, which consists of all integrations of the propagators, and a colour coefficient~$C(D)$, which includes all colour matrices and structure constants. In the case of untraced Wilson lines, the colour indices are thus all included in the colour coefficient. The kinematical parts are the same in QCD and in the replicated theory, so all dependence on~$N$ will appear in the colour coefficient only. We will use~$C_N(D)$ to distinguish the colour coefficient in the replicated theory from the colour coefficient~$C(D)$ in QCD.

These colour coefficients are determined in the following way: One attaches a replica index~$i$ to each gluon, writes down all colour matrices and structure constants which appear in the diagram according to the new path ordering in the replicated theory, and then sums each replica index from~$1$ to~$N$. Interacting gluons are required to have the same replica index. We can expand each colour factor in~$N$ and according to equation~\eqref{replica} the linear term then gives the coefficient with which this diagram appears in the exponent.

The colour coefficients are tensors with a number of indices which is twice the number of Wilson lines in~$W$. The number of independent parameters of such a tensor is in general given by the product of the maximal values of each index, but because the colour coefficients are calculated only with colour matrices in various representations, their number of independent parameters is usually much smaller. We can write them as a linear combination of a certain number of basic tensors, which are determined by the number and colour representation of the Wilson lines in~$W$.

We will use the Polyakov loop correlator as an illustrating example. We can write
\begin{align}
 P_c(\mathbf{r})&=\frac{1}{N^2_c}\left\langle\mathrm{Tr}\bigl[P(\mathbf{r})\bigr]\mathrm{Tr}\left[P^\dagger(0)\right]\right\rangle\notag\\
 &=\frac{\delta_{ji}\delta_{lk}}{N^2_c}\left\langle P_{ij}(\mathbf{r})P_{kl}^\dagger(0)\right\rangle\,,
\end{align}
where $i$, $k$ are final and $j$, $l$ are initial indices. The colour matrices~$T^a$ that will appear in the Feynman diagrams are all in the fundamental representation and all of their octet indices~$a$ are contracted. Because fundamental colour matrices with contracted octet indices can be decomposed in terms of Kronecker deltas and also structure constants can be expressed through fundamental colour matrices, there remain only two possibilities in which the initial and final indices can be combined in fundamental tensors:
\begin{equation}
 (t_1)_{ik,jl}=\delta_{ij}\delta_{kl}\,,\hspace{40pt}(t_2)_{ik,jl}=\delta_{il}\delta_{kj}\,.
\end{equation}

For example the colour coefficient of the one-gluon exchange diagram between the two Polyakov loops, which we will call $D_I$, is given by
\begin{equation}
 C(D_I)=T^a_{ij}T^a_{kl}=T_F\left(\delta_{il}\delta_{kj}-\frac{1}{N_c}\delta_{ij}\delta_{kl}\right)=-\frac{T_F}{N_c}t_1+T_Ft_2\,.
\end{equation}
In this case the standard colour factor $C(D_I)$ is identical to the exponentiated colour factor $\widetilde{C}(D_I)$, but in general they are different.

\begin{figure}[t]
 \centering
 \includegraphics[width=\linewidth]{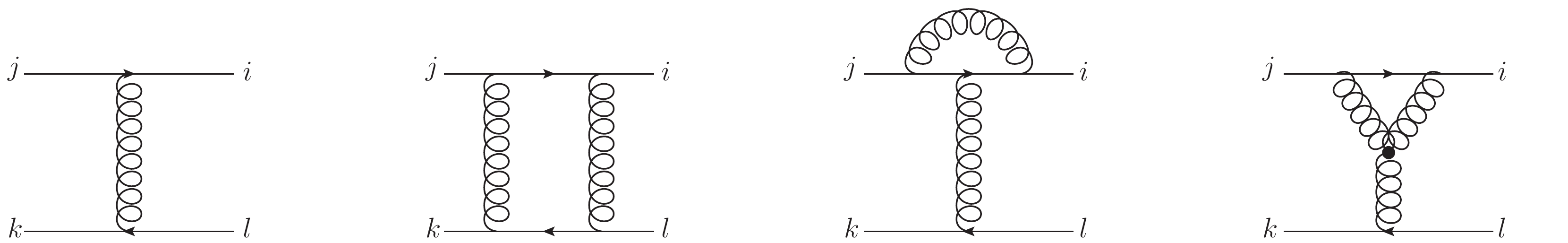}
 \caption{All one- and two-gluon exchange diagrams contributing to the exponentiated Polyakov loop correlator. The imaginary time axis runs horizontally, the vertical axis is in the direction of~$\mathbf{r}$. The propagators can be understood as dressed, because inserting loops does not affect their colour factors. The last two diagrams turned upside down contribute, but are not shown here explicitly.}
 \label{2gluonPc}
\end{figure}

In the tensor space of~$t_1$ and~$t_2$, $t_1$ is the unit element, and~$t_2$ has the property $t^2_2=t_1$. The exponential of a linear combination of~$t_1$ and~$t_2$ is therefore given by
\begin{equation}
 \exp\bigl[At_1+Bt_2\bigr]=\exp[A]\,\bigl(\cosh[B]\,t_1+\sinh[B]\,t_2\bigr)\,.
\end{equation}
Because diagrams without any gluons connecting the two Polyakov loops will always have a colour factor proportional to~$t_1$ only, we can factorize them out. They give the square of a single Polyakov loop and we then only have to concern ourselves with diagrams where one or more gluons are exchanged between the Polyakov loops. Figure~\ref{2gluonPc} shows all diagrams involving one or two exchanged gluons whose exponentiated colour factors do not vanish. We will call them $D_I$, $D_{II}$, $D_T$ and $D_Y$ from left to right. With $\delta_{ji}\delta_{lk}(t_1)_{ik,jl}=N^2_c$ and $\delta_{ji}\delta_{lk}(t_2)_{ik,jl}=N_c$ we get
\begin{align}
 P_c(\mathbf{r})&=\frac{1}{N^2_c}\bigl\langle\mathrm{Tr}\bigl[P(0)\bigr]\bigr\rangle^2\,\exp[A]\left(\cosh[B]+\frac{1}{N_c}\sinh[B]\right)\,,\label{ExpPc}\\
 A&=-\frac{T_F}{N_c}F(D_I)-T^2_F\bigl(F(D_{II})-2F(D_T)+2iF(D_Y)\bigr)+\dots\,,\\
 B&=T_FF(D_I)+T^2_FN_c\bigl(F(D_{II})-2F(D_T)+2iF(D_Y)\bigr)+\dots\,.
\end{align}

As an example of the calculation of the exponentiated colour factors, we will consider the diagrams~$D_{II}$ and the corresponding diagram~$D_X$ where the exchanged gluons are crossed. For two exchanged gluons there are~$N(N-1)$ possibilities to attach different replica indices to them and~$N$ possibilities to attach the same replica index. If the two gluons in~$D_{II}$ have a different replica index, then the replica path ordering requires that the colour matrices on one of the Polyakov loops be reversed, so we get a contribution equal to~$C(D_X)$. Then we have
\begin{align}
 C_N(D_{II})&=N(N-1)C(D_X)+NC(D_{II})=N\bigl(C(D_{II})-C(D_X)\bigr)+{\cal O}(N^2)\,,\\
 C_N(D_{X})&=N(N-1)C(D_X)+NC(D_{X})={\cal O}(N^2)\,,\\
 \widetilde{C}(D_{II})&=C(D_{II})-C(D_X)=T^2_F\left(\delta_{il}\delta_{mn}-\frac{1}{N_c}\delta_{im}\delta_{nl}\right)\left(\delta_{mn}\delta_{kj}-\frac{1}{N_c}\delta_{mj}\delta_{kn}\right)\notag\\
 &-T^2_F\left(\delta_{in}\delta_{km}-\frac{1}{N_c}\delta_{im}\delta_{kn}\right)\left(\delta_{ml}\delta_{nj}-\frac{1}{N_c}\delta_{mj}\delta_{nl}\right)=T^2_FN_c\left(t_2-\frac{1}{N_c}t_1\right)\,,\\
 \widetilde{C}(D_X)&=0\,.
\end{align}

Note that diagrams like~$D_I$, $D_T$ and~$D_Y$ do not appear in a direct perturbative calculation of the Polyakov loop correlator but only in the exponentiation formula~\eqref{ExpPc}, because their traced colour factor is zero. It can be checked by reexpanding the exponentiated expression that they only start to contribute at quadratic order.

In the case of $W_c-P_c$ the situation is more complicated and a simple exponentiated expression cannot be obtained, because there are many more fundamental tensors involved. But still, for the treatment of the linear divergences this generalized exponentiation formula is very useful. According to the classification established in~\cite{Ren1}, linear divergences can only arise from diagrams where all gluons are attached to the same Wilson line. Strictly speaking, this statement is gauge dependent, it does not apply to singular gauges such as axial gauges. But since we are ultimately dealing with gauge invariant quantities, we are free to choose a gauge where it holds true, such as covariant gauges or the Coulomb gauge.

Now, such diagrams with all gluons attached to the same Wilson line will always be proportional to the unit tensor, irrespective of which colour representation that Wilson line is in~\cite{Ren1}. All terms proportional to the unit tensor will appear simply as an exponential multiplying the rest of the loop function; compare the term~$A$ in the previous example.

Linear subdivergences can appear if a diagram has a subdiagram, but then the exponentiated colour factor is zero~\cite{Exp3,Exp4}. By the term \textit{subdiagram} we mean the following: if it is possible to cut the same Wilson line twice such that the cut out part is not connected to any other part of the diagram through gluons, then this part is called a subdiagram. We exclude from this definition the trivial cases when no gluons are attached to the cut out part at all or when it is the same as the whole diagram. A subdiagram is also proportional to the unit tensor only, so the colour factor of the whole diagram is given by the product of the colour factors of the subdiagram and of the rest of the diagram.
\begin{equation}
 C(D)=C(D_{\mathrm{sub}})\cdot C(D_{\mathrm{rest}})\,.
\end{equation}

This statement remains true even if by replica path ordering the subdiagram is split into smaller subdiagrams. The combinatorial factors from distributing replica indices to gluons can also be factorized: the total number of distributions of indices to all gluons is given by the number of distributions to the subdiagram times the number of distributions to the rest of the diagram. Therefore we can also write
\begin{equation}
 C_N(D)=C_N(D_{\mathrm{sub}})\cdot C_N(D_{\mathrm{rest}})\,.
\end{equation}
Both colour coefficients~$C_N(D_{\mathrm{sub}})$ and~$C_N(D_{\mathrm{rest}})$ are at least linear in~$N$, so the colour coefficient of the whole diagram is at least of quadratic order. Therefore such a diagram does not contribute to the exponent.

Ultimately this means that also in the case of untraced Wilson lines in general colour representations the linear divergences will always appear in the form of an exponential of a linearly divergent constant times the length of the respective Wilson lines. In general the linearly divergent constants, generically denoted~$\Lambda_R$, will depend only on the colour representation~$R$ and on the renormalization scheme used.

Finally, we have that
\begin{equation}
 \exp\bigl[-2\Lambda_F\beta-\Lambda_Ar\bigr]\,\times\,Z_{\,W_c-P_c}\,\times\,\bigl(W_c(\mathbf{r})-P_c(\mathbf{r})\bigr)\label{RenWP}
\end{equation}
is a finite quantity, where now~$Z_{\,W_c-P_c}$ is understood in the same renormalization scheme as the linear divergences. Equation~\eqref{RenWP} provides the renormalized expression of $W_c-P_c$ suited
for lattice calculations.

We expect that a similar relation holds in general for loop functions with overlapping Wilson lines. The overlapping parts can be decomposed into a linear combination of single Wilson lines in various colour representations in the same way that a direct product of irreducible $SU(N_c)$ representations can be decomposed into a direct sum. Then each term in the decomposition of the overlapping Wilson lines gets a linearly divergent factor~$\exp\left[\Lambda_RL(C)\right]$ according to its representation. We expect that they can be removed by a linearly divergent renormalization matrix mixing the associated loop functions that correspond to different path orderings at the endpoints of the overlapping parts. In addition, there will be intersection divergences at these endpoints.

If we compare the renormalization constant for the intersection divergences of the cyclic Wilson loop with the cusp renormalization constant of a rectangular Wilson loop, we see that $Z_{\,W_c-P_c}$ at leading order in~$\alpha_s$ is equal to the renormalization constant of an adjoint Wilson loop with two cusps of angle~$\pi/2$. At low orders in~$\alpha_s$, loop functions depend on the colour representation only through the quadratic Casimir $C_R$, which appears as an overall coefficient. At higher orders this so-called Casimir scaling no longer holds, so it would be an interesting subject for further study to see, whether the relation between~$Z_{\,W_c-P_c}$ and the adjoint cusp renormalization constant still holds beyond the breakdown of Casimir scaling.

\acknowledgments
The first part of this work has been done while M.~B.\ was at Kyoto University. M.~B.\ thanks the Nuclear Theory Group and in particular Prof.~Hideo Suganuma for their warm hospitality and interesting discussions, and the GCOE for financial support. M.~B.\ thanks Prof.~Hideo Suganuma for suggesting the investigation of a vacuum loop function consisting of two rectangular Wilson loops that lie next to each other so that one side overlaps, which led to the results of section~\ref{sec1}. We acknowledge financial support from the DFG cluster of excellence \textit{Origin and Structure of the Universe} (www.universe-cluster.de).

\end{document}